\documentstyle[psfig,prb,aps,twocolumn]{revtex}
\draft
\begin{document}

\title{Quantum internal modes of solitons in 1d easy-plane antiferromagnet \\
in strong magnetic field}
\author{B. A. Ivanov and A. K. Kolezhuk}
\address{Institute of Magnetism, National Academy of Sciences
and Ministry of Education of Ukraine,\\
36 Vernadskii av., 252142 Kiev,
Ukraine
}

\date{\today}
\maketitle

\begin{abstract}
In presence of a strong external magnetic field the dynamics of
solitons in a one-dimensional easy-plane Heisenberg
antiferromagnet exhibits a number of peculiarities. 
Dynamics of internal soliton degrees of freedom is 
essentially quantum, and they are strongly coupled to the ``translational''
mode of soliton movement.
These peculiarities lead to considerable changes in
the response functions of the system
which can be detected experimentally.
\end{abstract}


\maketitle

\narrowtext

\section{Introduction}
\label{sec:intro}

Study of soliton excitations in one-dimensional (1d) systems have
been attracting much attention since the pioneering work by
Krumhansl and Schrieffer. \cite{krumh75} 
Mikeska \cite{mikeska78} was the first one who calculated the 
contribution of solitons into observable quantities for the realistic system,
viz. quasi-1d easy-plane ferromagnet CsNiF$_3$ in external
magnetic field. His prediction of the 
soliton-induced central peak in the dynamical structure factor
was confirmed experimentally,
\cite{kjems&steiner78} which gave rise to intensive theoretical and
experimental studies. Since then,
soliton signatures were reliably observed in static and dynamic
characteristics of various 1d magnets with both ferro- and
antiferromagnetic coupling, see recent reviews.
\cite{mikeska&steiner91,ivkol95rev}

Easy-plane 1d antiferromagnetic compound TMMC ((CD$_3$)$_4$NMnCl$_3$) in
external field is one of the most widely studied materials
\cite{mikeska80,boucher&renard80,harada&81,fluggen&mikeska83,%
regnault&82,boucher&84,gouvea&pires86,wysin&86,boucher&86} (see
also Ref.\ \onlinecite{mikeska&steiner91} and references therein). 
Its magnetic ions have spin $S=5/2$, and thus it is believed that the dynamics
of TMMC can be well described on the basis of classical equations of motion
for the sublattice magnetization. It is well
known that in presence of a strong magnetic field the equations of
motion for antiferromagnet are rather complicated and cannot be
reduced to the sine-Gordon (SG) model because of the importance of out-of-plane
deviations. \cite{fluggen&mikeska83,wysin&86} 
Effective field-induced anisotropy competes with the easy-plane anisotropy,
so that at certain critical field $H_c\simeq (H_e H_a)^{1/2}$  (where $H_e$
and $H_a$ are the exchange and anisotropy fields, respectively)  in-plane and
out-of-plane magnon modes have equal excitation energies, and the SG
approximation becomes obviously inadequate for $H$ close to $H_c$.

At $H=H_c$ the ``easy plane'' effectively changes.  There are, respectively,
two types of static kink solutions, \cite{harada&81} one of which has lower
energy at $H<H_c$ and the other at $H>H_c$. Corresponding dynamic solutions
are much less understood. Different authors
\cite{fluggen&mikeska83,wysin&86,costa&pires85,iko91,ivanov&kolezhuk91}
obtained contradictory results concerning their stability.  It was shown
\cite{wysin&86,iko91,ivanov&kolezhuk91} that the effective plane of rotation
of the sublattice magnetization in a kink depends on the kink velocity, and
the two types of kinks transform into each other at some critical velocity
$v_c$, $v_c\to0$ at $H\to H_c$.

The main feature of the problem, determining kink stability and dynamics in
strong magnetic fields, is the presence of out-of plane magnon mode localized
at the kink. \cite{iko91,ivanov&kolezhuk91,ivkol95rev} This internal mode is
highly nonlinear at $v\to v_{c}$ and strongly coupled to the ``translational''
(zero) mode of the kink motion, \cite{ivkol95rev} which makes its analysis
rather complicated. Moreover, it can be shown that because of the relatively
small effective mass of solitons [the ratio of soliton and magnon masses is
$S$ for antiferromagnet, while for ferromagnets this ratio is proportional to
the large parameter $(H_{e}/H_{a})^{1/2}$] the dynamics of internal soliton
degrees of freedom in antiferromagnets is essentially quantum even for
``almost classical'' values of $S$ like $5/2$ \cite{ivkol95rev,ivkol95prl} and
exhibits subtle effects distinguishing between integer and half-integer $S$.
\cite{ivkol95fnt,ivkol96jetp} Thus, to our opinion, the internal dynamics of
solitons in TMMC in a strong-field regime can be highly nontrivial, and its
consistent description, as well as experimental observation, is a fascinating
problem.

In this paper we present a detailed theoretical study of the 
soliton dynamics in 1d ``almost-classical'' spin-$S$ easy-plane
antiferromagnet in strong external magnetic field close to the
critical value $H_c$.  We show that quantum effects (tunneling between two
energetically equivalent kink states) lead to appearance of a new 
localized mode which is also coupled to the translational mode and thus 
depends on the soliton momentum $P$. At $P=0$ the tunneling is suppressed for
half-integer $S$ due to the topological effects, but at nonzero $P$ it becomes
possible for any $S$. We calculate the contribution of the localized modes
into the dynamic structure factor (DSF) and show that it leads to an additional
peak concentrated at nonzero frequency, which can be
detected by electron spin
resonance (ESR) or inelastic neutron scattering (INS) technique. Numerical
estimations of the corresponding resonance frequencies and the peak properties
for TMMC are given.
To the knowledge of authors, at present the only known experiment probing
internal soliton modes is the so-called  ``soliton
magnetic resonance'' \cite{boucher&87} in the Ising-type quasi-1d
antiferromagnet 
CsCoCl$_3$, and no such data exist for the Heisenberg
magnets.

The paper is organized as follows: In Section \ref{sec:model} we introduce the
model. In Sect.\ \ref{sec:internal-classic} we
derive an effective model of internal kink dynamics, perform classical linear
analysis of the spectrum of excitations against the background of kinks and
discuss the soliton stability problem, and in Sect.\ 
\ref{sec:internal-quantum} we perform quantization of the effective model and
discuss the behavior of the energy levels of localized modes.  In Sect.\ 
\ref{sec:dsf} the soliton contribution to the response functions is
considered, and numerical estimations of predicted effects for TMMC are given.
Finally, Sec.\ \ref{sec:summary} contains concluding remarks.

\section{Model}
\label{sec:model}

We consider a one-dimensional two-sublattice Heisenberg antiferromagnet (AFM)
of the TMMC type in external magnetic field. This system is well
described by the following Hamiltonian:
\begin{eqnarray}
{\cal H} &=&\sum_n\big\{ J\vec{S}_n\vec{S}_{n+1} \nonumber\\
&+& D_1 (S_n^z)^2
- D_2 (S_n^y)^2 -g\mu_B\vec{H}\vec{S}_n\big\}\;,
\label{ham}
\end{eqnarray}
where $J>0$ is the exchange constant, $\mu_B$ is the Bohr magneton, $g$ is the
Lande factor, $D_1$ and $D_2$ are anisotropy constants, $D_1\gg D_2>0$,
$\vec{H}$ is the magnetic field, and spins $\vec{S}_n$ are treated as
classical vectors. In absence of the field $(xy)$ is an easy plane, and $y$ is
the easiest axis in this plane (the case of ``pure'' easy-plane AFM
corresponds to $D_2=0$); the chain axis is along $z$. For TMMC the value of
the spin is $S={5\over2}$, and the following values for the parameters of the
classical Hamiltonian have been established: \cite{regnault&82} $JS^2=85$~K,
$D_1S^2=1.9$~K, $D_2S^2=0.022$~K, and $g=2.01$. 

Long-wavelength dynamics of AFM can be described within the
nonlinear sigma model approach.
It is convenient to introduce the
antiferromagnetism vector (the sublattice magnetization)
$\vec{l}_i=(\vec{S}_{2i+1}-\vec{S}_{2i})/2S$ at each magnetic elementary cell.
At low temperatures, when the magnetization
$\vec{m}_i=(\vec{S}_{2i+1}+\vec{S}_{2i})/2S$ is small, $\vec{l}$ can be
regarded as a unit vector, and after passing to the continuum limit the
following Lagrangian can be obtained:
\begin{eqnarray}
\label{lagr}
 L &=& {1\over2}JS^{2}a \int dz \left\{ {1\over c^2} (\partial_t \vec{l})^2 
- (\partial_z \vec{l})^2  
-{1\over x_{0}^{2}}l_z^2 -{1\over x_{b}^{2}}\right. \nonumber\\
  &-& \left. {\gamma^{2}\over c^{2}}(\vec{l}\cdot\vec{H})^2 
 + {2\gamma\over c^{2}}
  \vec{H}\cdot(\vec{l}\times\partial_t\vec{l})\right\} \\
 &+& {\hbar S\over 2}\int dz \vec{l}\cdot(\nabla\vec{l}\times\partial_t\vec{l})
\;.\nonumber
\end{eqnarray}
Its derivation from different points of view the reader
can find in Refs.\ \onlinecite{ibarbob79-80,andrmar80,affleck85,haldane85};
this approach is closely related to the commonly used Mikeska's
formulation. \cite{mikeska80} Here $a$ is the lattice spacing,
$c=2JSa/\hbar$ is the limiting velocity of spin waves,
$\gamma=g\mu_{B/\hbar}$ is the gyromagnetic ratio,  $x_{0}=
a[J/2(D_{1}+D_{2})]^{1/2}$ and $x_{b}=a(J/2D_{2})^{{1/2}}$ are the
characteristic length scales. 
The last term in (\ref{lagr}) is the so-called topological term which
can be rewritten as $2\pi\hbar S Q$, where
$Q$ is the Pontryagin index (the winding number) of a given space-time
configuration of the field $\vec{l}$. This term is irrelevant in classical
analysis, but is very 
important for the quantum theory since it distinguishes between integer and
half-integer $S$.

Within this approximation, the vector of magnetization $\vec{m}$ is a
``slave'' variable and can be expressed through the antiferromagnetism vector
$\vec{l}$:
\begin{equation}
\vec{m}={\hbar\over4JS}(\vec{l}\times\partial_t\vec{l})
+{g\mu_{B}\over4J}\{\vec{H}-\vec{l}(\vec{H}\cdot\vec{l})\}
-{1\over2}a\nabla\vec{l} \;.
\label{m}
\end{equation}

Further, it is convenient to introduce angular variables for the
vector $\vec{l}$. We choose the easiest axis $y$ as a polar one,
and put
\[
l_y=\cos\theta\;,\quad l_z=\sin\theta\cos\varphi\;,\quad
l_x=\sin\theta\sin\varphi\;.
\]
Let us set the field in plane perpendicular to the easy axis,
$\vec{H}\parallel\widehat{\vec{x}}$, then
$y$ is always the easiest axis, and there is no spin-flop
transition. 
However, at $H=H_c$, where
\begin{equation}
H_c=2(2JD_{1}S^{2})^{1/2}\;,
\label{Hc}
\end{equation}
the axes $x$ and $z$ switch their roles: the easy plane is $xy$
when $H<H_c$, and $yz$ when $H>H_c$. In this paper, we consider the case
of strong fields $H\approx H_{c}$; for TMMC
the critical field is of the order of 100~kOe. 
In what follows, we can safely
neglect any contribution coming from $D_{2}$ because in strong fields
the field-induced anisotropy in the basal plane is much
larger than the initial crystal-field anisotropy.
Hereafter, we assume that the field is close to
$H_{c}$, and regard the quantity
\begin{equation}
  \label{rho}
  \rho=1-(H/H_{c})^{2}
\end{equation}
as a small parameter playing the role of effective rhombicity (as far as only
static properties are considered, $H=H_{c}$ corresponds to a ``uniaxial''
situation).

\section{Effective model of kink dynamics}
\label{sec:internal-classic}

Because of the presence of magnetic field, the equations of motion for the
model (\ref{lagr}) contain not only the usual Lorentz-invariant terms, but
also terms of the first order in time derivatives, and
thus their analysis is rather complicated. However, one can easily find static
soliton solutions. There are two types of static kinks, with $\vec{l}$
rotating in the $(xy)$ plane ($\varphi=\pm\pi/2$) or in the $(yz)$ plane
($\varphi=0,\,\pi$). \cite{harada&81} The behavior of the angle $\theta$ in
both kinks is given by
\begin{equation}
\cos\theta_0=\pm\tanh(z-z_{s}/z_0)\;,
\label{kinktheta}
\end{equation}
where the kink thickness is $z_0=x_{0}$ for the $(yz)$-kink and $z_{0}\simeq
x_{0}H_{c}/H=x_{0}/(1-\rho)^{1/2}$ for the $(xy)$-kink, and $z_{s}$ is an
(arbitrary) kink coordinate.
For $H<H_c$ the kink with $\varphi=\pm\pi/2$
is the lowest energy nonlinear excitation, and for $H>H_c$ this
role goes to the kink with $\varphi=0,\,\pi$; at $H=H_c$ their
energies are equal, as well as the energies of in-plane and
out-of-plane magnon branches. 

As it was shown in Refs. \onlinecite{gomonai&90,iko91}, approximate soliton
solutions can be constructed using the following trick: One can easily see
that the characteristic scale of $\varphi$ variation
$l_\varphi=x_{0}/sqrt{\rho}$ is much larger than the characteristic thickness
of the kink $x_0$. Then, ``within the kink,'' i.e., within the region where
the variation of $\theta$ mainly takes place, we can neglect spatial
dependence of $\varphi$ and put $\varphi=\varphi_{s}=\text{const}$.

We use the same trick, but regard $z_{s}$ and $\varphi_{s}$ as new dynamic
variables slowly varying in time: we assume that the soliton solution has the
form
\begin{eqnarray}
&& \cos\theta_{s}=\sigma\tanh\left( {z-z_s(t) \over z_{0}(\varphi_{s})}
\right)\;,\quad \varphi=\varphi_s(t)\;, \nonumber\\
&& z_{0}(\varphi)\equiv x_{0}(1-\rho\sin^{2}\varphi)^{-1/2} ,
\label{ansatz}
\end{eqnarray}
here $\sigma=\pm1$ is the topological charge distinguishing kinks and
antikinks. Substituting the ansatz (\ref{ansatz}) into the Lagrangian
(\ref{lagr}) and assuming that $z_{s}$ and $\varphi_{s}$ are ``slow''
variables, i.e.
\begin{equation}
  \label{slow}
  \dot{z}_{s} \ll c\,, \qquad 
 \dot{\varphi}_{s} \ll \omega_{0}\equiv c/x_{0}\,,
\end{equation}
one can obtain the following effective Lagrangian with only two degrees of
freedom:
\begin{eqnarray}
  L_{\text{eff}}&=&-E_{0}(1-\rho\sin^{2}\varphi_s)^{1/2} +{E_{0}\over2c^{2}}
  (\dot{z}_{s}^{2} +x_{0}^{2}\dot{z}_{s}^{2})\nonumber\\
  &+&{\pi E_{0}\over 2c}\sigma \dot{z}_{s}(1-\rho)^{1/2}\cos\varphi_{s}-\hbar
  S\dot{\varphi}_{s}  \,.
\label{Leff}
\end{eqnarray}
Here the dot denotes differentiation with respect to time, and
$E_{0}=2S^{2}(2JD_{1})^{1/2}$ is a characteristic energy scale (rest energy of
a soliton at $H=H_{c}$). The last term in the Lagrangian comes from the
topological term in (\ref{lagr}) and is ineffective in classical treatment
since it is a complete derivative. However, in quantum treatment it changes
the definition of canonical momentum conjugate to $\varphi_{s}$ which is very
important, as we will see below.

The effective potential for the internal variable $\varphi_{s}$ strongly
depends on the kink velocity $\dot{z}_{s}$. One can see that a classical
analysis of the effective Lagrangian (\ref{Leff}) allows one to reproduce the
results of a more rigorous approach of Refs.\ 
\onlinecite{iko91,ivanov&kolezhuk91,ivkol95rev} which we briefly revisit
below. The canonical momentum $P$ conjugate to $z_{s}$,
\begin{equation}
  \label{P}
  P={E_{0}\over c^{2}}\dot{z}_{s}+P_{0}(1-\rho)^{1/2} \sigma
  \cos\varphi_{s}\,, \quad\text{where\ } P_{0}={\pi E_{0}\over 2c}\,,
\end{equation}
is conserved, and thus it is convenient to express everything in terms of $P$.
The effective potential
\begin{eqnarray}
 \label{Ueff}
 U_{\text{eff}}&=&E_{0}(1-\rho\sin^{2}\varphi_{s})^{1/2}\\
  &+&{\pi^{2}E_{0}\over8} \left\{
    {P\over P_{0}}- \sigma\cos\varphi_{s}(1-\rho)^{1/2} \right\}^{2}\nonumber
\end{eqnarray}
has four extrema as a function of $\varphi_{s}$, at
$\varphi_{s}=0,\pi$ and at $\varphi_{s}=\varphi_{1,2}(P)$, where
$\varphi_{1,2}(P)$ are two roots of the equation
\begin{eqnarray}
\label{xy-kink}
&&  \cos(\varphi_{s})=\sigma P/P_{c}\;,\\
&& P_{c}\simeq P_{0} \left\{ 1-\rho\left({1\over2}-{4\over \pi^{2}}
  \right) \right\}\,.\nonumber
\end{eqnarray}
Extrema at $\varphi_{s}=0,\pi$ correspond to $(yz)$-kinks, and extrema at
$\varphi_{s}=\varphi_{1,2}(P)$ correspond to $(xy)$-type kinks. When the 
velocity of an $(xy)$-type kink is zero, the plane of rotation of the vector
$\vec{l}$ is $(xy)$, and 
this plane smoothly changes to $(yz)$ when $P$ tends to $P_{c}$, so that
$(xy)$-type kink smoothly transforms into a $(yz)$-kink.
The frequency of linear
oscillations around those extrema is given by
\begin{equation}
  \label{omega-loc}
  \omega^{2}_{l}=\omega_{0}^{2}f(\rho) 
  \cases{ \left(\sigma_{0} \displaystyle{P\over
        P_{c}}-1\right) & for 
    $\varphi_{s}=0,\pi$\cr 
  \left(1-\displaystyle{P^{2}\over P^{2}_{c}}\right) & for
    $\varphi_{s}=\varphi_{1,2}(P)$ \cr} , 
\end{equation}
where
\begin{eqnarray*}
 && \sigma_{0}=\sigma\,\text{sgn}(\cos\varphi_{s})\,,\\ 
 && f(\rho)\simeq{\pi^{2}\over 4} \left\{ 1-{\pi^{2}-4\over\pi^{2}} \rho
 \right\}\,. 
\end{eqnarray*}
Thus, $(xy)$-type kinks are stable in the entire region $-P_{c}\leq P \leq
P_{c}$ where the corresponding solutions (\ref{xy-kink}) have sense. The
stability of $(yz)$-kinks depends on $P$ and on the sign of the
quantity $\sigma_{0}=\text{sgn}(\theta_{0}'\cos\varphi_{s})$: $(yz)$-kinks
with $\sigma_{0}>0$ are stable at $P>P_{c}$, and those with $\sigma_{0}<0$ are
stable at $P<-P_{c}$. For each $(yz)$-kink there exist two possible
``daughter'' $(xy)$-type kinks, corresponding to the two possible solutions
$\varphi_{1,2}(P)$, so that the transition from $(yz)$-kink to $(xy)$-kink when
$|P|$ drops below $P_{c}$ is a spontaneous breaking of symmetry.

The dispersion relation for soliton excitations, i.e. the dependennce of their
energy $E$ on the momentum $P$, is schematically shown in Fig.\ 
\ref{fig:levels} (the curve labeled ``classical'').  For $\rho>0$, i.e.
$H<H_{c}$ the minimal energy is reached at $P=0$, while at $\rho<0$
($H>H_{c}$) the minimum is located at $P=P_{0}$; in both cases the second
derivative $d^{2}E/dP^{2}$ is discontinuous at $P=\pm P_{c}$.

It should be remarked that the effective model (\ref{Leff}) was derived under
the assumption that the variable $\varphi_{s}$ is ``slow,'' and therefore
expressions (\ref{omega-loc}) have sense only if $\omega_{l}^{2}\ll
\omega_{0}^{2}$. This does not matter, however, for the classical analysis of
stability presented above, because the corresponding frequencies are small in
the vicinity of the transition. But what does matter is (i) strong
nonlinearity: in linear analysis the localized mode frequency is zero at
$P=P_{c}$, which indicates that anharmonic terms have to be taken into
account; (ii) quantum effects: the magnitude of quantum zero-point
fluctuations around the classical minima of the effective potential strongly
increases when the potential well becomes ``flat.''

\section{Quantum internal modes}
\label{sec:internal-quantum}

Since the  effective model (\ref{Leff}) contains
only two degrees of freedom, one can easily quantize it in a canonical way.
The Lagrangian does not depend on $z_{s}$ and thus the total
momentum $P$ is conserved and can be treated as a $c$-number, so that we are
left with only one variable $\varphi_{s}$. 
The Hamiltonian can be easily obtained in the form
\begin{equation}
  \label{Heff}
  \widehat H = {\hbar^{2}\omega_{0}^{2}\over 2E_{0}}\left\{
    i{\partial\over\partial\varphi_{s}} +S \right\}^{2} +U_{\text{eff}}(\varphi_{s}) 
 \,,
\end{equation}
where $S$ in curly brackets appears because the last term in (\ref{lagr})
changes the definition of the angular momentum. As we will see below, the
presence of this term changes the situation drastically depending on whether
$S$ is integer or half-integer. For the wavefunction $\Psi(\varphi_{s})$ one
has usual periodic boundary conditions $\Psi(\varphi)=\Psi(\varphi+2\pi)$.

The effective potential $U_{\text{eff}}$ is generally two-well, with
energetically 
equivalent well minima; at $P=0$ the minima are located at
$\varphi_{s}=\pm\pi/2$, as $P$ deviates from zero they both move towards $0$
or $\pi$ depending on the sign of $P$ and the `generalized' topological charge
$\sigma_{0}$, and at $P=\pm P_{c}$ they merge into one well at $\varphi_{s}=0$
or $\pi$. One may expect that tunneling between equivalent minima at
$|P|<P_{c}$ splits the lowest energy levels, leading to the appearance of one
more internal soliton mode. However, using arguments of the same type as
in Refs.\ \onlinecite{ivkol95fnt,ivkol96jetp},  it is easy to show that at
$P=0$ the 
tunneling is possible only for integer $S$. Indeed, the tunneling amplitude is
proportional to the path integral
\begin{equation}
  \label{path}
  \int _{\varphi_{1}(P)}^{\varphi_{2}(P)}
  {\cal D}\varphi_{s}\exp\left\{{i\over\hbar}\int \,dt L_{\text{eff}}
  \right\}\,, 
\end{equation}
where $\varphi_{1,2}(P)$ are the well minima and the effective Lagrangian is
defined by Eq.\ (\ref{Leff}). Since $L_{\text{eff}}$ contains the topological
term $\hbar S\dot{\varphi_{s}}$, different paths contribute with the phase
factor proportional to $e^{iS\Delta\varphi_{s}}$, where $\Delta\varphi_{s}$ is
the change in $\varphi_{s}$ along the path. At $P=0$ one has two equivalent
paths with $\Delta\varphi_{s}=\pm\pi$, so that their contributions cancel each
other for half-integer $S$ (recall that in case of TMMC $S={5\over2}$). If
$P\not=0$, the two paths become inequivalent, thus allowing nonzero tunneling
amplitude. 

\begin{figure}
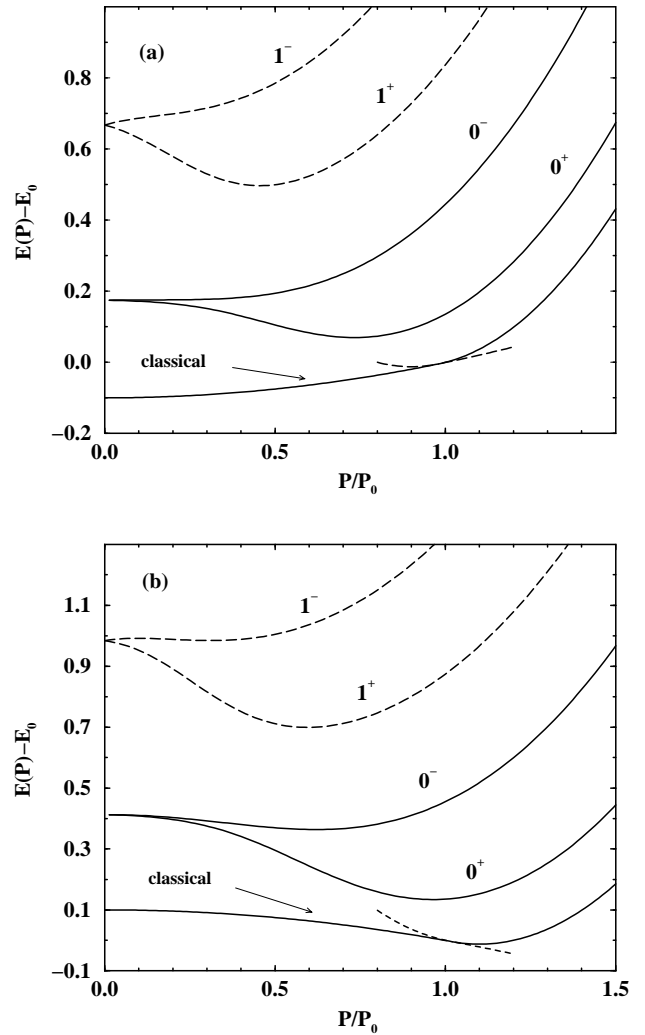

\caption{
  Typical dependence of the soliton energy levels $E$ on its momentum $P$ for
  half-integer spin $S$ (levels presented in this figure were computed for
  $S={5\over2}$): (a) $H<H_{c}$ ($\rho=0.2$); (b) $H>H_{c}$ ($\rho=-0.2$).
  The curve labeled ``classical'' corresponds to the  result of classical
  calculation 
  \protect\cite{ivkol95rev} for the lowest energy state; dashed pieces of the
  curve indicate unstable states. For $S={5\over2}$ the $1^{\pm}$ levels are
  completely inside the magnon continuum and thus are also shown with dashed
  lines.  For integer $S$ the degeneracy at $P=0$ is lifted.
\label{fig:levels}  
}
\end{figure}

The Schr\"odinger equation with the Hamiltonian (\ref{Heff}) can be easily
solved numerically, and one gets the energy eigenvalues as functions of the
soliton momentum $P$; a typical level structure is shown in Fig.\
\ref{fig:levels}. All states at fixed $P$
\[
\psi_{n,\nu}(P,\varphi_{s})=u_{n,\nu}(P,\varphi_{s})e^{iS\varphi_{s}}
\]
can be labeled with two quantum numbers $n$ and $\nu$ ($\nu$ being the parity
of the ``reduced'' wavefunction $u$), and for half-integer $S$ odd and even
states are degenerate at $P=0$. One can see that quantum effects change the
soliton dispersion considerably comparing to the classical calculation;
particularly, the dispersion tends to have a minimum at nonzero $P$ even for
fields lower than $H_{c}$, 
\vskip -0.1in
\mbox{\hspace{-0.3in}\psfig{figure=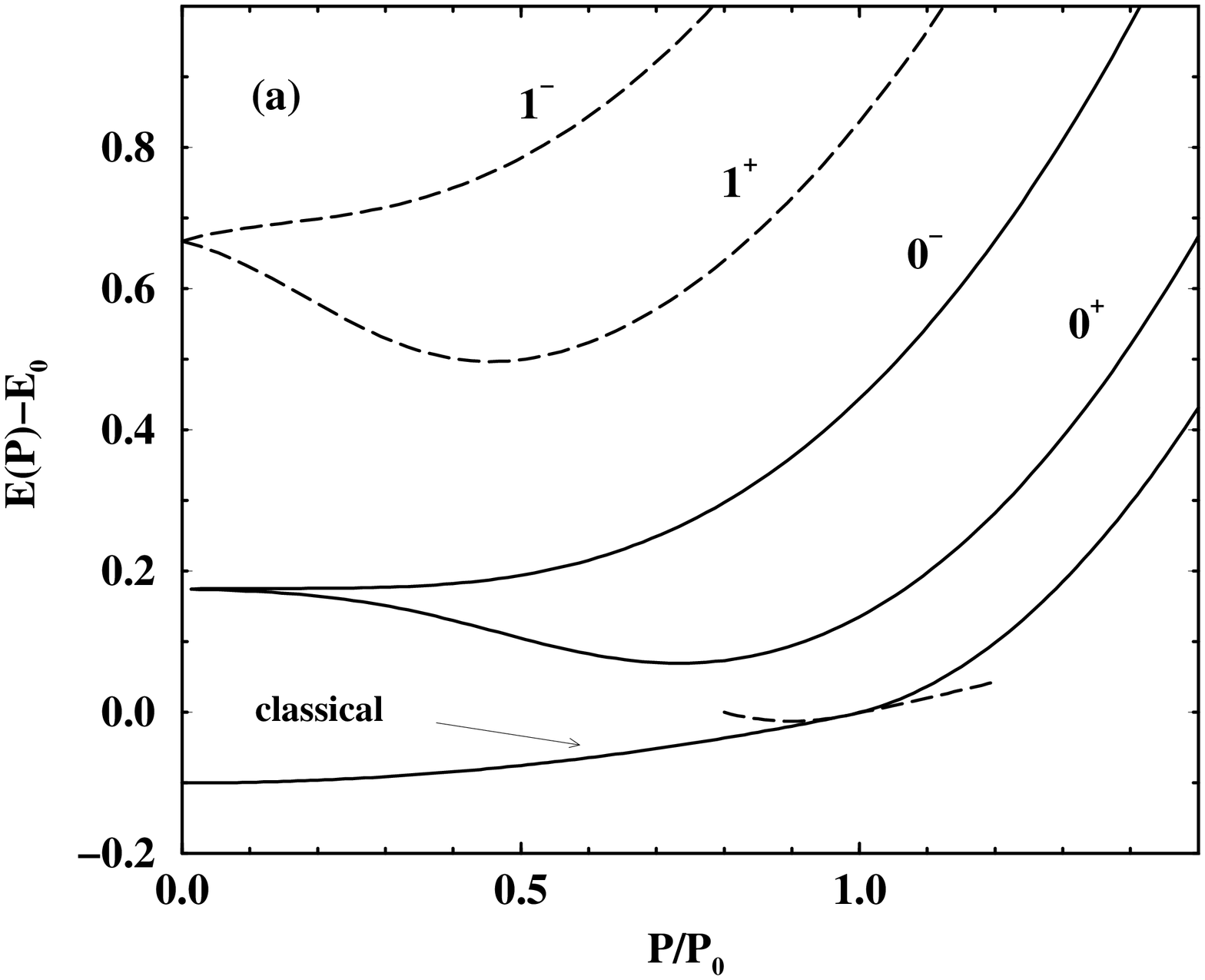,height=3in,angle=-90.}}
 \vskip -0.2in\nopagebreak
\mbox{\hspace{-0.3in}\psfig{figure=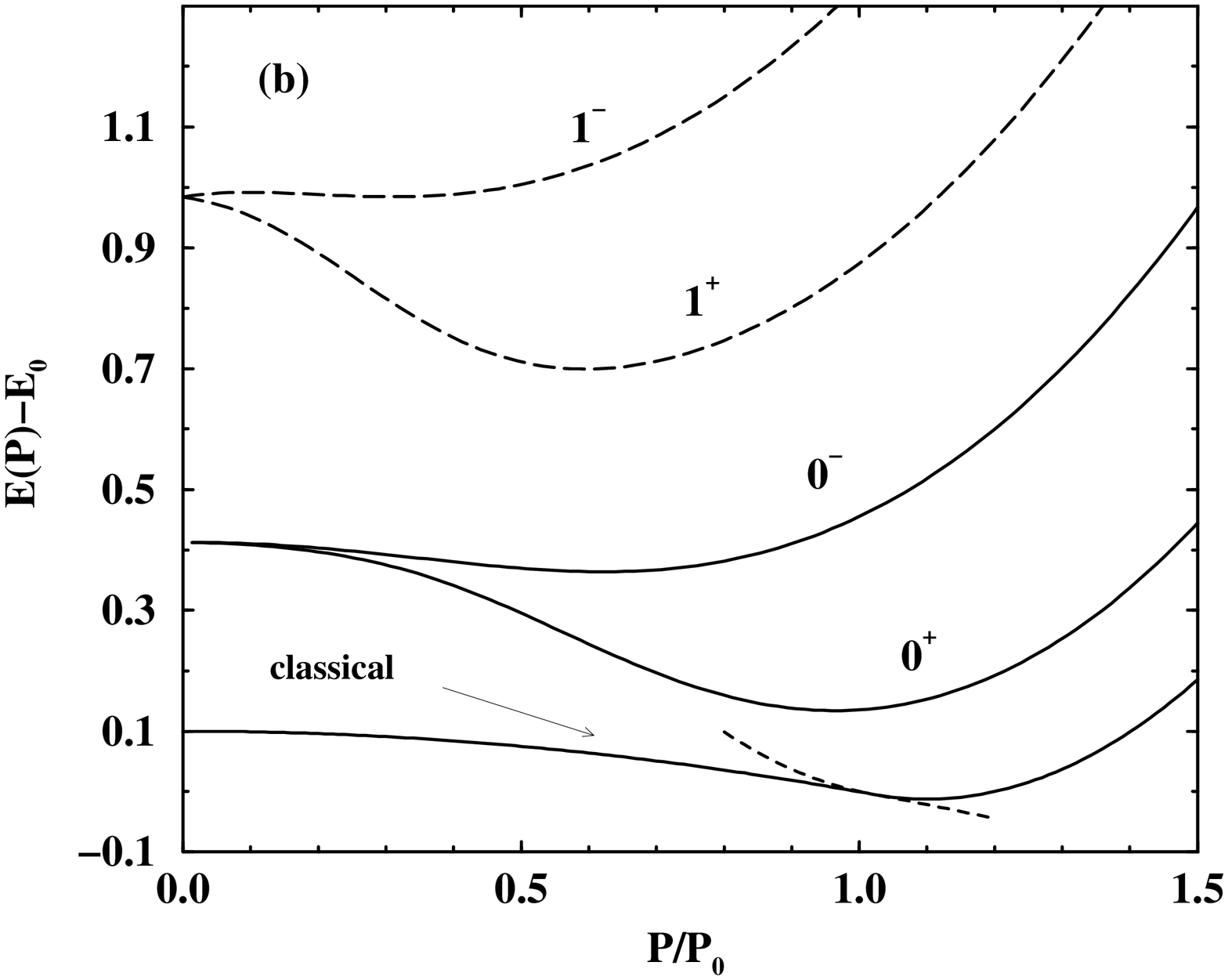,height=3in,angle=-90.}}
  \noindent\parbox[t]{3.40in}{\protect\small
FIG.\ \ref{fig:levels}. 
Typical dependence of the soliton energy levels $E$ on its momentum $P$ for
  half-integer spin $S$ (levels presented in this figure were computed for
  $S={5\over2}$); (a) $H<H_{c}$ ($\rho=0.2$); (b) $H>H_{c}$ ($\rho=-0.2$).
  The curve labeled ``classical'' corresponds to the result of classical 
  calculation
  \protect\cite{ivkol95rev} for the lowest energy state; dashed pieces of the
  curve indicate unstable states. For $S={5\over2}$ the $1^{\pm}$ levels are
  completely inside the magnon continuum and thus are also shown with dashed
  lines.  For integer $S$ the degeneracy at $P=0$ is lifted.
}
\vskip 0.1in\noindent
which means that $H_{c}$ is effectively
renormalized towards its decrease. This renormalization is not small, as seen
from Fig.\ \ref{fig:levels}, since at $\rho=0.2$ the lowest energy is still
reached at $P\not=0$. Unfortunately, we are not able to calculate this
renormalization quantitatively:
first of all, our effective model is valid only at small $\rho$, and second,
there are other quantum effects influencing the effective soliton energy (the
change in the energy of zero-point fluctuations of magnon modes in presence of
a kink) which are not considered here; the interested reader is referred for
details to the original papers \cite{dashen&75,maki81,mikeska82} and to the
recent review. \cite{mikeska&steiner91} The lowest excitations of internal
soliton mode clearly have purely quantum nature, i.e., they do not have any
classical counterpart, and are determined by the tunnel level splitting.

\section{Internal soliton modes and response functions}
\label{sec:dsf}

In this section we apply the results obtained so far to find out how 
the excitation of internal soliton degrees of freedom 
contributes into the response functions of easy-plane antiferromagnet in 
a strong-field regime ($H$ close to $H_c$).

We wish to begin with a remark concerning the relation between
the gap in the spectrum of linear excitations $\hbar\omega_0$ and
the kink rest energy $E_0$ in various soliton-bearing systems.
For the model of structural phase transition used by Krumhansl
and Schrieffer $\hbar\omega_0\propto \Phi_0$, and $E_0\propto
\Phi_0^3$, where $\Phi_0$ is the magnitude of the order
parameter, so that near the transition point
$E_0\ll\hbar\omega_0$, which justifies neglecting kink-phonon
interaction at $T\ll E_0$ in Ref.\ \onlinecite{krumh75}. For ferromagnets one
obtains $E_0/\hbar\omega_0\simeq (J/D)^{1/2}$, where $J$ and $D$
are the exchange and anisotropy constants, respectively. In case of
the Heisenberg magnets $J/D\gg1$, and $E_0\gg \hbar\omega_0$, so that
the inequality $\hbar\omega_0\ll T\ll E_0$, usually assumed in
soliton-gas calculations, can be easily satisfied. For
antiferromagnets the situation is ``intermediate'':
$E_0/\hbar\omega_0\simeq S$, and the
inequality $E_0/\hbar\omega_0$ holds only for $S\gg1$. The
relation $S\gg 1$ is often used as a condition under which a
quantum spin system can be treated classically, but in actual
practice the strong inequality is not needed, and classical
equations of motion work well, say, for $S=5/2$, as in case of
TMMC. Thus, in antiferromagnets for realistic $S$ values $E_0$ is
only a few times greater than $\hbar \omega_0$, and usually used
condition $\hbar\omega_0\ll T\ll E_0$ is not valid as a strong inequality. (For
example, in case of strong fields close to $H_c$
the characteristic values of kink energy and
magnon gap for TMMC are $E_0\simeq20$~K and
$\hbar\omega_0\simeq7$~K, respectively).

To justify the phenomenological approach \cite{krumh75,currie&80} to the
soliton thermodynamics, the only necessary condition is smallness of the kink
gas density $n_0$ comparing to the inverse kink thickness $x_0^{-1}$. This
means smallness of the exponent $\exp(-E_0/T)$, i.e., practically, it is
sufficient that $T\preceq {1\over3}E_0$. Therefore, for antiferromagnets
instead of a strong inequality $\hbar\omega_0\ll T$ one must consider the
converse condition $\hbar\omega_0\succeq T$.

The most general quantity determining the response functions is the so-called
dynamical structure factor (DSF) $S^{\alpha\beta}(q,\omega)$, which is
essentially the space-time Fourier component of the two-spin correlation
function $\langle S^\alpha(n,t)S^\beta(n',0)\rangle$.  In case of
antiferromagnet, the quantities of interest are (i) the DSF component at
$q=0$, describing the intensity of response in electron spin resonance (ESR)
experiments; (ii) the DSF component at $q\approx Q_{B}$, where $Q_{B}$ is the
Bragg wavevector, which is usually measured in inelastic neutron scattering
(INS) experiments. The main contribution to the DSF at $q=Q_{B}$ comes from
the correlator of the antiferromagnetism vector
$\langle\vec{l}(z,t)\vec{l}(z',0)$, and the component at $q=0$ is determined
by the correlator of magnetization
$\langle\vec{m}(z,t)\vec{m}(z',0)$. \cite{sasaki84} 
Soliton contribution to the correlation functions can be calculated using the
approximation of the soliton ideal gas,\cite{krumh75} which means that at low
density of kinks the main role is played by one-soliton correlations. Then one
has to deal with correlators of the form
\begin{eqnarray}
&& \int dz\int dz' e^{iq(z-z')}\nonumber\\
&& \qquad \times\langle f\big(z-z_s(t)\big) A(\varphi_s,t)
f\big(z'-z_s(0)\big) A(\varphi_s,0) 
\rangle\,, 
\label{corr}
\end{eqnarray}
where $A$ and $f$ are generally certain operators. The average should be
taken in both quantum-mechanical and thermodynamic senses. 
Quantum-mechanical averaging is performed with the wavefunctions
$\Psi_{\lambda,P}=e^{iPz_{s}}\psi_{\lambda}(P,\varphi_{s})$ (we use the
notation $\lambda=\{n,\nu\}$ for the sake of brevity), and  the
thermodynamic average should be performed with the Gibbs distribution function
\begin{equation}
w_\lambda(P)=(2\pi\hbar)^{-1}
e^{-E_{\lambda}(P)/T}e^{-\Sigma_{s}(\lambda,P)/T}\,,
\label{w}
\end{equation}
where $w$ is normalized to the total density of solitons, and
the quantity $\Sigma_{s}$ is the change in
the free energy of the magnon gas due to the presence of a kink,
which is attributed to the kink in the phenomenological
approach: \cite{currie&80}
\FL
\[
\Sigma_s(\lambda,P)=T\sum_j\int dk\,\Delta n^{(j)}(k,\lambda,P)
\ln[2\sinh(\hbar\omega^{(j)}_k/T)]\;.
\]
Here $j=1,2$ labels different branches of magnons of the continuous
spectrum, and $\omega^{(j)}_k$ are the frequencies of
corresponding modes. The change in the magnon density of states
$\Delta n^{(j)}(k,\lambda,P)$ is connected with the phase shift
$\Delta_s^{(j)}(k,\lambda,P)$ acquired by a magnon of the $j$-th branch with
wavevector $k$ after its interaction
with the kink having quantum numbers $\lambda$, $P$: 
$\Delta n^{(j)}=(1/2\pi)(d\Delta_s^{(j)}/dk)$.
[In fact, $\Sigma_s$ contains divergent terms
connected with the change in the energy of zero-point
vibrations between the ground state and the one-soliton state.
This divergence, however, is compensated by the counterterms
arising from the normal ordering of operators,
yielding finite quantum correction to the
soliton energy. \cite{dashen&75,maki81,mikeska82}]

Using the above arguments,  expression (\ref{corr}) can be rewritten as
\begin{eqnarray}
&& |f(q)|^{2}\sum_{\lambda,\lambda'}\int dP e^{-(E_{\lambda}(P)
+\Sigma_s(\lambda,P))/T} 
\big| \langle \lambda,P|\widehat A
|\lambda',P-\hbar q\rangle\big|^{2} \nonumber\\
&&\qquad\qquad\qquad
 \times\,\delta(E_{\lambda}(P)-E_{\lambda'}(P-\hbar q)-\hbar\omega)\,,
  \label{aver}
\end{eqnarray}
where $|\lambda,P\rangle\equiv \psi_{\lambda}(P)$. At low temperatures $T\ll
E_{0}$ one may take into account only the contribution of the lowest two
soliton levels $0^{\pm}$ (see Fig.\ \ref{fig:levels}).

At $T\ll\hbar\omega_0$, $\Sigma_s$ is proportional to
$\exp(-\hbar\omega_0/T)$ and can be neglected.  At $T\succeq\hbar\omega_0$ the
effect of $\Sigma_s^{\mbox{cont}}$ becomes significant; its accurate
calculation requires solving the soliton-magnon scattering problem to get the
phase shifts, which is too complicated to be done analytically.  However,
$zy$-type solitons, i.e.  kinks with $P\geq P_{c}$, classically have
$\varphi_{s}=0,\pi$, thus being ``sine-Gordon'' in structure, and the phase
shifts for the sine-Gordon (SG) problem are known.  In this case
$\Delta_{s}^{(1,2)}(k)\simeq -2\arctan(kx_0)$ at small kink velocities $v\ll
c$, and for $T\gg\hbar\omega_0$ we obtain the well-known SG result
\[
\Sigma_s=-2T\ln2-2T\ln(\hbar\omega_0/T)\;.
\]
On the other hand, the same SG result should be valid for $xy$-solitons with
$P=0$.  We will assume that the same expression can be applied for all $P$;
this assumption seems plausible though it cannot be justified in a rigorous
way. One can expect that possible weak dependence of $\Sigma_{s}$ will lead to
certain renormalization of the $P$-dependence of the soliton energy levels
$E_{\lambda}$, but we believe that  our results will remain qualitatively
correct. 

\subsection{Internal modes in ESR}

The static component of magnetization $\vec{m}$, according to (\ref{m}), is
directed along the $Ox$ axis. Let us assume that the pumping rf field is
parallel to $Oz$, then the intensity of the ESR signal is proportional to
$S^{zz}(0,\omega)$ and is determined by the correlator of $m_{z}$.  It is easy
to see that only the first term in $m_{z}$, according to (\ref{m}),
contributes to $S^{zz}(0,\omega)$ at nonzero frequencies [the last term in
(\ref{m}) does not depend on $\varphi_{s}$ and thus yields the delta function
in $\omega$, and the contribution from the second term just vanishes because
at $q=0$ the corresponding formfactor $f(q)$ in (\ref{aver}) is zero]. Thus,
the significant term in $m_{z}$ is proportional to
$\sin^{2}\theta_{s}\dot{\varphi}_{s}$, and $\dot{\varphi}_{s}$, according to
(\ref{Leff}) is proportional to $\widehat p_{\varphi}-\hbar S$, where
$\widehat p_{\varphi}=-i\hbar{\partial\over\partial\varphi_{s}}$ is the
operator of the canonical momentum conjugate to $\varphi_{s}$. Again, $\hbar
S$ yields $\delta(\omega)$, so that the only contribution at nonzero $\omega$
comes from $p_{\varphi}$.  Typical plots of the ESR response depending on
frequency for $S={5\over2}$ are shown in Fig.\ \ref{fig:esr}; one can see that
the transitions between $0^{+}$ and $0^{-}$ soliton levels lead to a
pronounced peak with a maximum located well below the antiferromagnetic
resonance frequency $\omega_{0}$.  Since the levels $0^{+}$ and $0^{-}$ have
different minima in their $P$ dependencies, the position of peaks in the ESR
response at positive and negative $\omega$ is also different.  When the
temperature increases, peaks shift towards $\omega_{0}$, but still keep their
shape with a clearly pronounced maximum.  Thus we expect that this effect can
be really observable experimentally in TMMC, where at $H=H_{c}$ the
three-dimensional ordering temperature $T_{N}\simeq3$K, i.e. about
$0.15E_{0}$.

\subsection{Internal modes in INS}

\begin{figure}
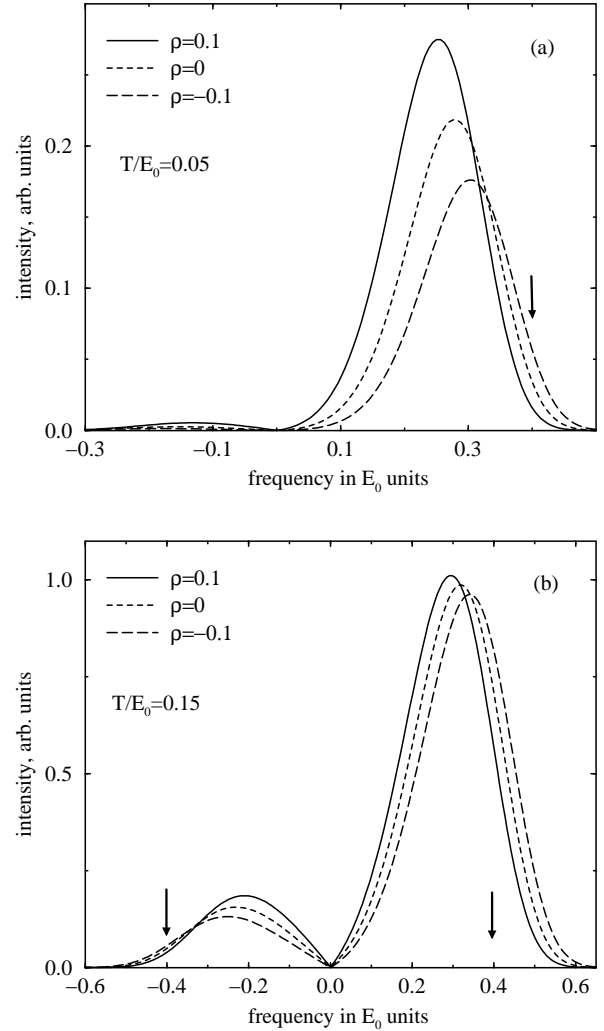

\caption{
\label{fig:esr}  
Plots of $S^{zz}(0,\omega)$ (this quantity determines the intensity of the ESR
signal) for $S={5\over2}$ and different values of the parameter
$\rho=1-H^{2}/H_{c}^{2}$:
(a) $T=0.05E_{0}$; (b) $T=0.15E_{0}$.  Arrows show the positions of the
boundary of the continuous magnon spectrum at $H=H_{c}$.
}
\end{figure}

\begin{figure}
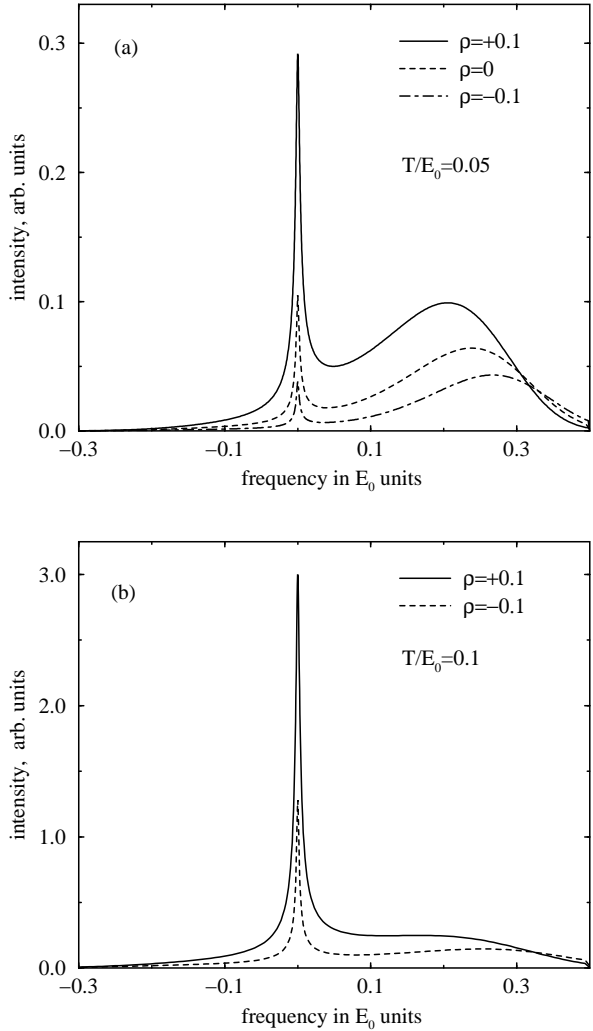

\caption{
\label{fig:dsf}
Plots of the longitudinal DSF $S^{xx}(Q_{B},\omega)$ for $S={5\over2}$ and
different values of the parameter 
$\rho=1-H^{2}/H_{c}^{2}$:
(a) $T=0.05E_{0}$; (b) $T=0.1E_{0}$.  
}
\end{figure}

In neutron experiments, it is usually possible to measure either longitudinal
(with respect to the magnetic field) DSF component
$S^{\parallel}=S^{xx}(q,\omega)$, or the transverse 
\vskip -0.1in
\mbox{\hspace{-0.3in}\psfig{figure=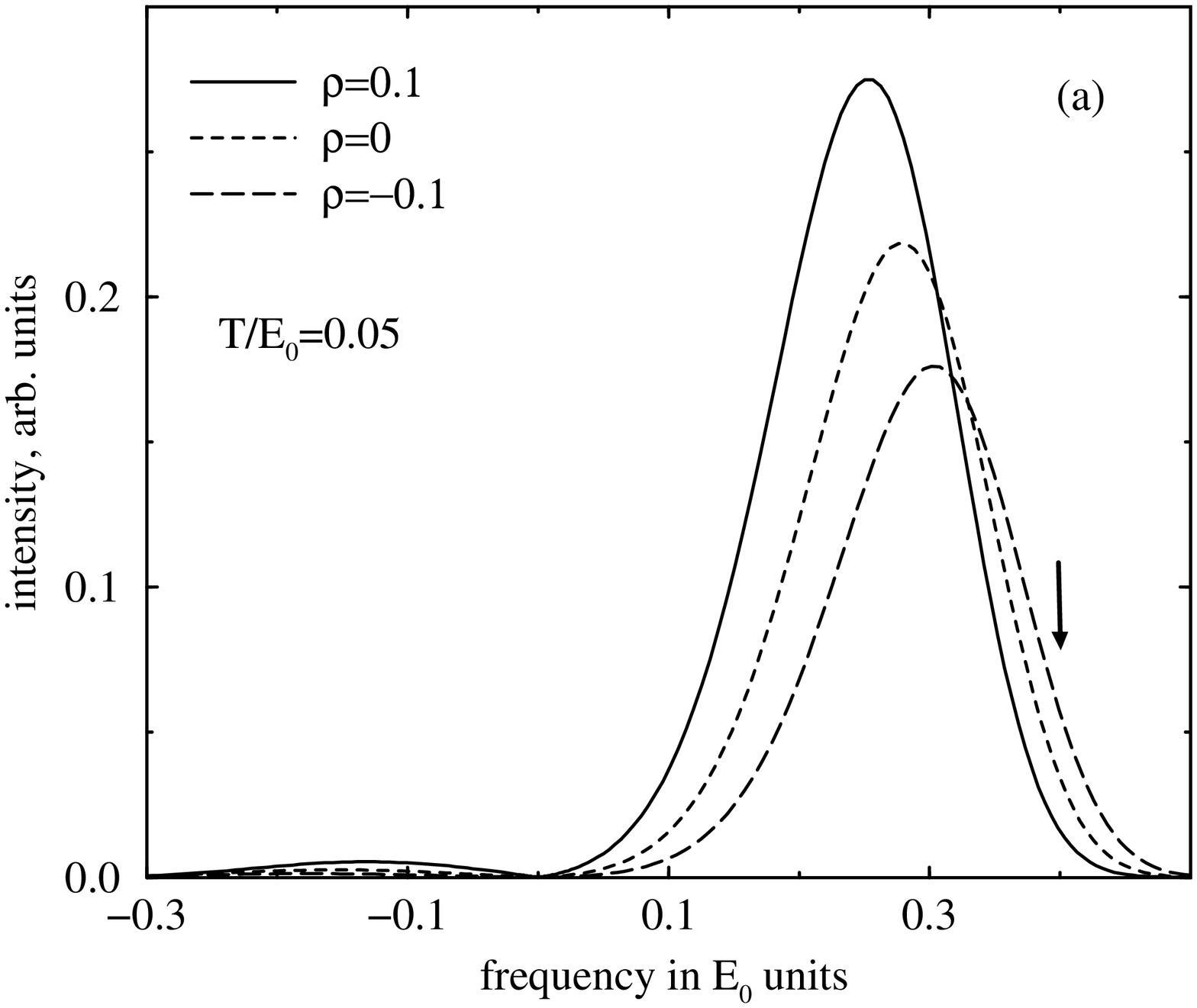,height=3in,angle=-90.}}
 \vskip -0.2in\nopagebreak
\mbox{\hspace{-0.3in}\psfig{figure=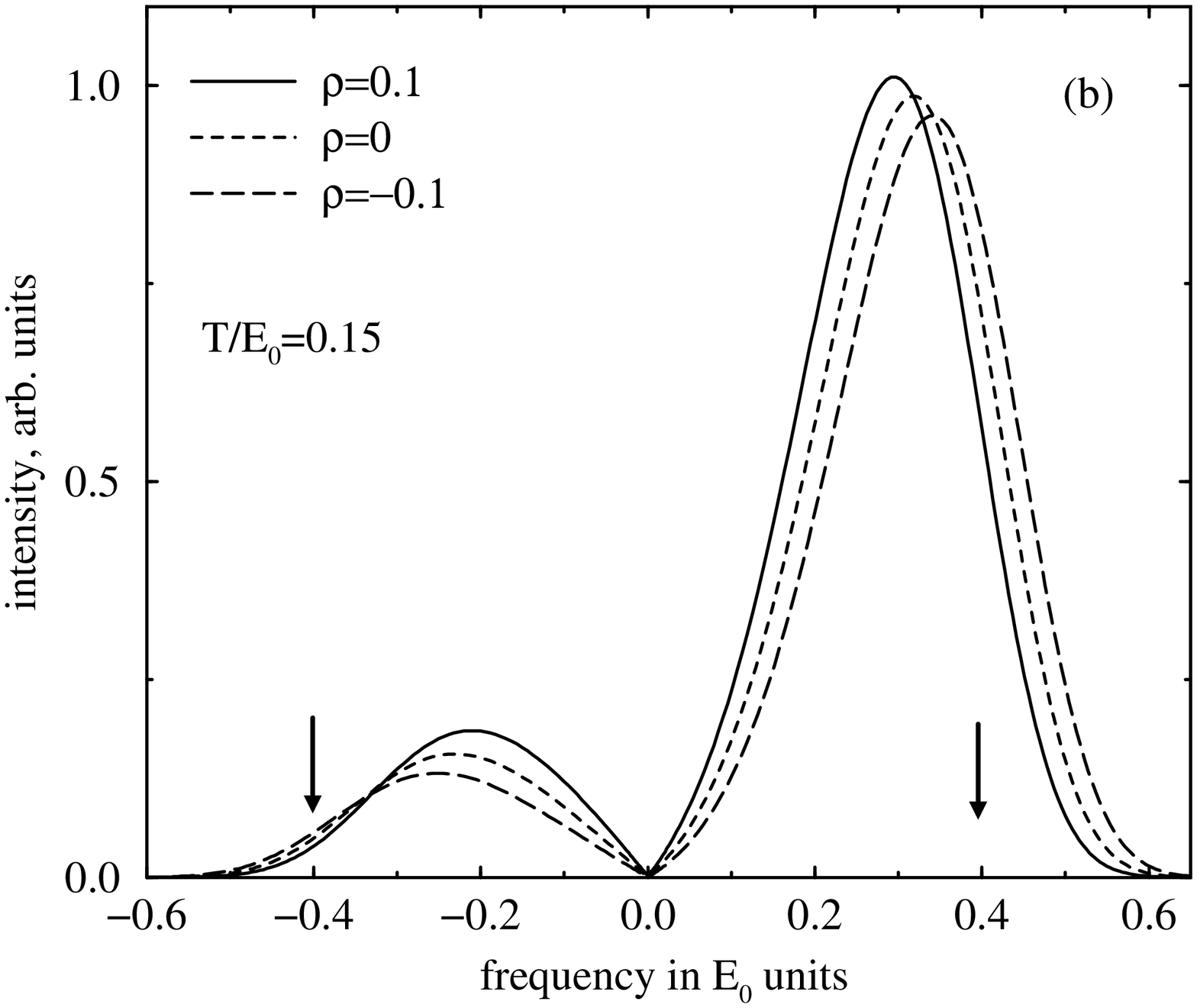,height=3in,angle=-90.}}
  \noindent\parbox[t]{3.40in}{\protect\small
FIG.\ \ref{fig:esr}. 
Plots of $S^{zz}(0,\omega)$ (this quantity determines the intensity of the ESR
signal) for $S={5\over2}$ and different values of the parameter
$\rho=1-H^{2}/H_{c}^{2}$:
(a) $T=0.05E_{0}$; (b) $T=0.15E_{0}$. Arrows show the positions of the
boundary of the continuous magnon spectrum at $H=H_{c}$.
}
\vskip 0.1in\noindent
one $S^{\perp}=S^{yy}(q,\omega)+S^{zz}(q,\omega)$, with $q$ close to the Bragg
wavevector $Q_{B}$. In TMMC, both those components exhibit presence of the
soliton central peak. \cite{boucher&86} In the transverse DSF the dominating
contribution comes from $S^{yy}$ 
(which corresponds to the contribution of flips of the direction
of the antiferromagnetism vector $\vec{l}$ along the easy axis $Oy$), and
$S^{yy}$ is insensitive to the effects of internal modes because $l_{y}$ does
not depend on $\varphi_{s}$; that is why we will be interested only in the
longitudinal component $S^{xx}$ determined by the correlator of $l_{x}$.
Fig.\ \ref{fig:dsf} shows several typical plots of $S^{xx}(Q_{B},\omega)$ for
different values of the field and temperature (actually, we have also
calculated $S^{xx}(q,\omega)$ for a few values of $q$ close to $Q_{B}$, but
the resulting curves exhibit very weak $q$ dependence, so that we restrict
ourselves to presenting only plots for $q=Q_{B}$). One can see that the most
significant manifestation of internal soliton modes is a nonzero width of the
central peak (CP) 
\vskip -0.1in
\mbox{\hspace{-0.3in}\psfig{figure=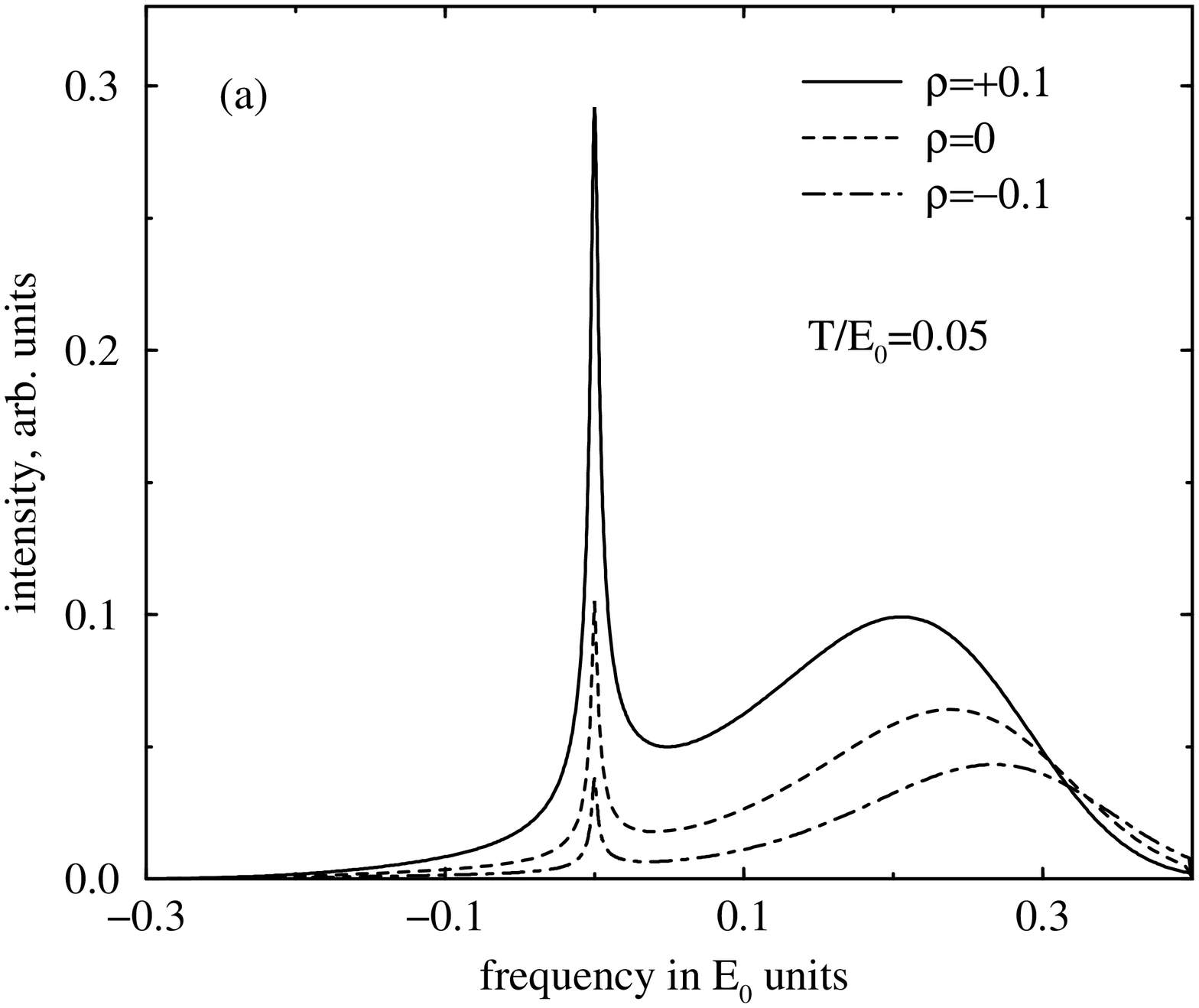,height=3in,angle=-90.}}
 \vskip -0.2in\nopagebreak
\mbox{\hspace{-0.3in}\psfig{figure=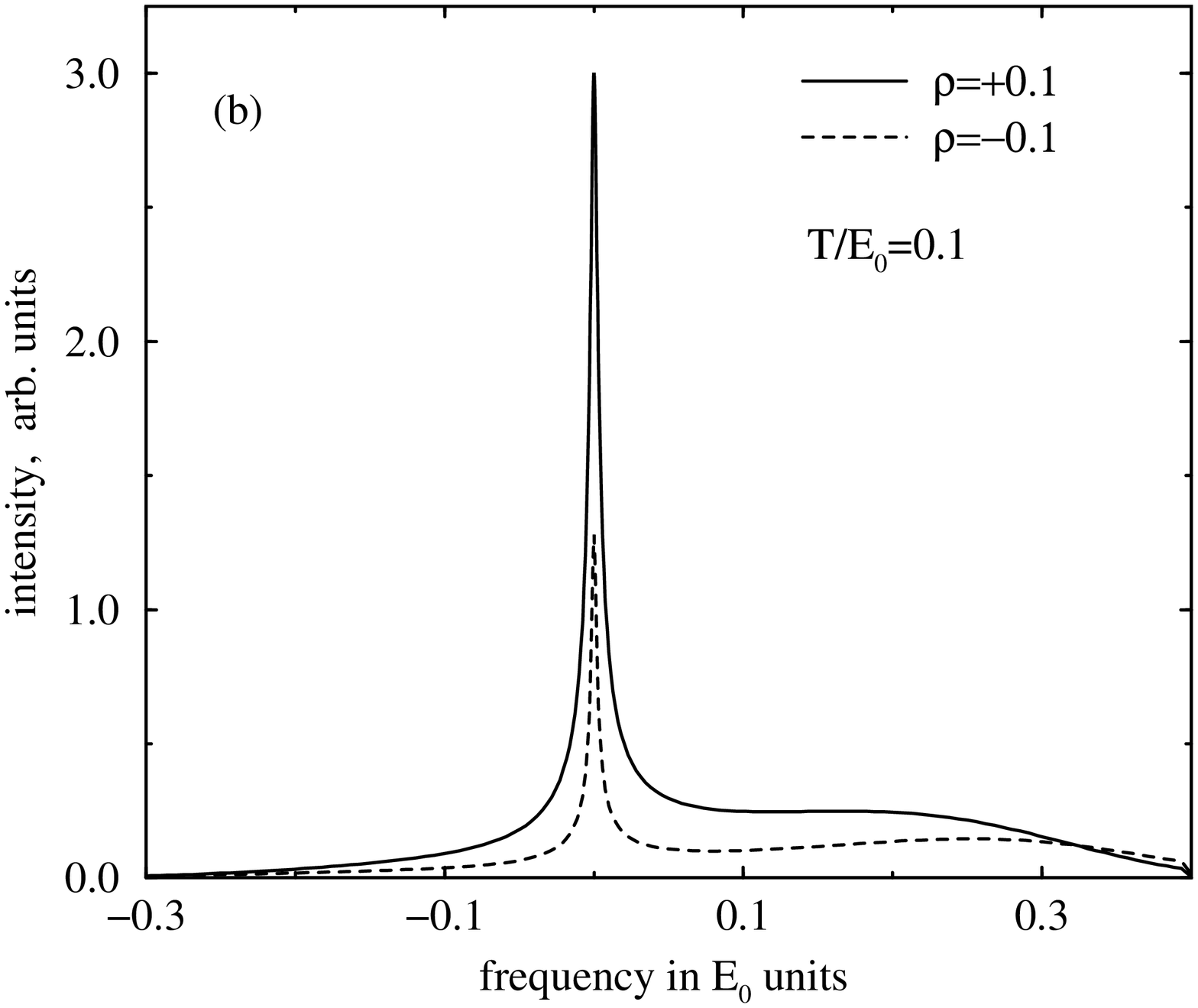,height=3in,angle=-90.}}
  \noindent\parbox[t]{3.40in}{\protect\small
FIG.\ \ref{fig:esr}. 
Plots of the longitudinal DSF $S^{xx}(Q_{B},\omega)$ for $S={5\over2}$ and
different values of the parameter 
$\rho=1-H^{2}/H_{c}^{2}$:
(a) $T=0.05E_{0}$; (b) $T=0.1E_{0}$.
}
\vskip 0.1in\noindent
at $q=Q_{B}$ (recall that in the standard ideal-gas calculation
\cite{mikeska80} without taking into account  internal degrees of freedom  the
CP width for the longitudinal component is proportional to $|q-Q_{B}|$).
However, it should be mentioned that the effects of soliton-magnon and
soliton-soliton scattering, which we do not consider here, also lead to
nonzero CP width at $q=Q_{B}$, \cite{SasakiMaki87} and therefore it may be
problematic to separate those two contributions experimentally. The additional
peak at nonzero frequency is also present in $S^{xx}$, but with increasing
temperature it quickly gets smeared by the tails of the dominating central
peak (see Fig.\ \ref{fig:dsf}b), and in TMMC at $H\approx H_{c}$ for $T\succeq
T_{N}\simeq0.15E_{0}$ this peak should be almost completely suppressed. 

\section{SUMMARY}
\label{sec:summary}

We consider the problem of soliton dynamics in 1d ``almost-classical''
spin-$S$ easy-plane antiferromagnet in strong external magnetic field, taking
into account internal degrees of freedom of solitons.  We show that quantum
effects of tunneling between two energetically equivalent kink states lead to
level splitting for all values of $S$; the usual selection rule\cite{} which
prohibits the tunnelling for half-integer $S$ is lifted due to the strong
coupling between internal and translational modes. It is predicted that this
internal mode can be detected in TMMC [(CD$_3$)$_4$NMnCl$_3$] by means of the
electron spin resonance (ESR) or inelastic neutron scattering (INS) technique;
we also show that ESR is a preferable method because the corresponding
resonance peak is much more pronounced comparing to that in the INS response.

\acknowledgements

The authors are grateful to V.~G.~Bar'yakhtar, H.-J.~Mikeska, and G.~M.~Wysin
for stimulating discussions.  This work was partially supported by the grant
2.4/27 ``Tunnel'' from the Ukrainian Ministry of Science and Technology.

\end{document}